\documentclass[twocolumn]{revtex4}
\usepackage[dvips]{graphicx}
\begin{document}

\title{
Large dielectric constant and giant nonlinear conduction in
the organic conductor $\theta-$(BEDT-TTF)$_2$CsZn(SCN)$_4$
}

\author{
K. Inagaki
}
\author{
I. Terasaki
}
\email{terra@waseda.jp}
\affiliation{
Department of Applied Physics, Waseda University, Tokyo 169-8555, Japan
}

\author{
H. Mori
}
\affiliation{
The Institute for Solid State Physics, The University of Tokyo, Kashiwa 277-8581, Japan\\
CERST, Japan Science and Technology Agency, Kawaguchi 332-0012, Japan
}

\author{
T. Mori
}
\affiliation{
Department of Organic and Polymeric Materials, 
Tokyo Institute of Technology, Tokyo 152-8552, Japan
}

\date{\today}

\begin{abstract}
The dielectric constant and ac conductivity have been measured for 
the layered organic conductor $\theta-$(BEDT-TTF)$_2$CsZn(SCN)$_4$ along the
out-of-plane direction, which show a relaxation behavior similar to
those in the charge-density-wave conductor.
Most unexpectedly, they exhibit a large bias dependence with a hysteresis,
and changes in magnitude by 100-1000 times at a threshold.
These findings are very similar to the collective excitation of 
the charge density wave.
$\theta-$(BEDT-TTF)$_2$CsZn(SCN)$_4$ 
has collective excitations associated with charge ordering,
though it shows no clear indication of long range order.
\end{abstract}

\pacs{}

\maketitle

\section{Introduction}
There appear growing interests in the self-organization 
phenomena of conduction electrons,
known as charge inhomogeneity.
A prime example is the high-temperature superconducting copper oxides,
in which the nano-scale phase-separation of doped carriers 
is observed in scanning-tunnel-microscope experiments \cite{davis}.
Another example is the charge order (CO)
in the transition-metal oxides \cite{cheong,tranquada},
where spins and holes are alternately
aligned to form a charge-density-wave (CDW) order
and a spin-density-wave (SDW) order simultaneously.
These have been regarded as a new aspect of strongly correlated electron systems,
because the electrons tend to be self-organized in order to minimize the 
magnetic energy of the background spins.
Thus we expect that such a self-organization will happen regardless of 
the degrees of disorder, and call it ``intrinsic inhomogeneity''.
In real materials, however, disorder-driven effects cannot be ruled out,
owing to unavoidable solid solutions and defects.
But rather, there increase a number of papers reporting that the 
charge order is quite susceptible to disorder \cite{ueda}.
Thus, to search for intrinsic charge inhomogeneity,
we need a charge-ordered system in a very clean solid.

Organic salts are a good candidate for the clean system.
They are grown in organic solvent at room temperature, where
the thermal energy is small enough to suppress point defects.
Actually most of the metallic salts exhibit 
the Shubnikov-de Haas oscillation at low temperatures \cite{review}.
This means that the mean free path of the conduction electron
is very long, and that the crystal is free from impurities.
In particular, the BEDT-TTF [bis(ethylenedithio)-tetrathiafulvalene] salts 
are most suitable for the study of the strong correlation in two dimension,
and the $\kappa-$(BEDT-TTF) salts have been most thoroughly investigated
among them, where a superconductor-antiferromagnetic insulator transition
takes place by changing hydrostatic pressure, anion species, 
and deuterium substitution \cite{kanoda,kanoda2}.

$\theta-$(BEDT-TTF)$_2M$Zn(SCN)$_4$ ($M=$Rb and Cs) is 
another interesting family of the BEDT-TTF salts \cite{mori,mori2},
where the BEDT-TTF molecules form a distorted triangular lattice.
The BEDT-TTF layer acts as a two-dimensional conducting layer,
and owing to the triangular stacking,
a single elliptic Fermi surface is calculated in the tight-binding approximation.
This makes a remarkable contrast with the $\kappa-$type salts,
in which the BEDT-TTF molecules form a dimer to make 
two Fermi surfaces (the electron pocket and the one-dimensional sheet).
In this sense, the $\theta-$type salts are a truly two-dimensional system
with a moderate in-plane anisotropy.
Another feature is that they are quarter-filled 
(one hole for the two BEDT-TTF molecules), which means that the system
is far from the Mott transition that occurs near half-filling \cite{kino}.
The most interesting feature is that 
they are unstable against CO along the $c^*$ axis \cite{seo}.
A metal-insulator transition due to CO occurs at $T_{\rm MI}=$190 K
for $\theta-$(BEDT-TTF)$_2$RbZn(SCN)$_4$ (the $M=$Rb salt) \cite{mori}.

The dielectric constant is a good probe for the charge inhomogeneity.
We have studied the dielectric response of 
strongly correlated systems, and have found a large
dielectric constant above $T_{\rm MI}$ for the $M=$Rb salt \cite{inagaki}.
The frequency dependence is quantitatively explained in terms of
a generalized formula of Debye's dielectric relaxation,
which strongly suggests that fractions of CO already
exist even at room temperature. 
Such a picture is consistent 
with other measurements such as NMR \cite{miyagawa} and 
optical reflectivity \cite{wang}.
This is a piece of evidence that the charge inhomogeneity can 
occur in a homogeneous system.
In this paper, we report on the dielectric response and 
the nonlinear conductivity of 
$\theta-$(BEDT-TTF)$_2$CsZn(SCN)$_4$ (the $M=$Cs salt) single crystals.
We have found anomalously large nonlinear conductivity 
just like the sliding of CDW,
although there is no sign for CDW/SDW transitions at all temperatures.
This implies that the $M=$Cs salt exhibits 
{\it a collective excitation without long range order}.

\section{Experimental}
Single crystals with a typical dimension of 
2$\times$0.2$\times$0.2 mm$^3$
were prepared using a galvanostatic
anodic oxidation method, and the detailed growth
conditions and their characterization were described in
\cite{mori}.

The resistivity was measured by a four-probe and two-probe method 
from 4.2 to 300 K in a liquid He cryostat.
The contact resistance was 10-50 $\Omega$ at room temperature.
Thus the resistivity obtained by the two-probe method
was accurate within an error bar of 1\%
at all the temperatures for the out-of-plane direction
(the $b$-axis direction),
and below 10 K for the in-plane direction (the $a$-axis direction).

The dielectric constant ($\varepsilon_b$) 
and ac conductivity ($\sigma_b$) along the out-of-plane direction
(the $b$ direction) were measured 
with a parallel plate capacitor arrangement 
using an ac two-contact four-probe method 
with an LCR meter (Agilent 4284A, and 4286A) 
from 10$^3$ to 10$^6$ Hz.
The dc-bias dependence of $\varepsilon_b$ and $\sigma_b$
was also measured in the static dc voltage.
The in-plane dielectric constant was not measured, because
it had too low resistance for us to measure $\varepsilon$ precisely.
Owing to the two-contact configuration,
the contact resistance and capacitance might have affected the measurement.
As mentioned, the contact resistance was negligible along the out-of-plane
direction, which was verified by the fact that
the observed $\sigma_{ac}$ was quantitatively consistent with
the observed dc resistivity for $\omega\to 0$.
Though we did not evaluate the contact capacitance, 
we can employ an evaluated value of 500 pF for La$_2$CuO$_4$ 
from Ref. \cite{chen},
because the contact capacitance is primarily determined 
by the area of the contact.
It gives a reactance of 3$\times$10$^5~\Omega$ at 1 kHz, 
which is 10$^4$ times larger than the contact resistance.
As a result, we can safely assume that our contact is primarily 
a resistive coupling, and can neglect the contact capacitance.
We should further note that we tested different contact configurations
of other samples, and found that
the results were reproducible within experimental errors.

\section{Results}

\begin{figure}[t]
 \begin{center}
  \includegraphics[width=8cm,clip]{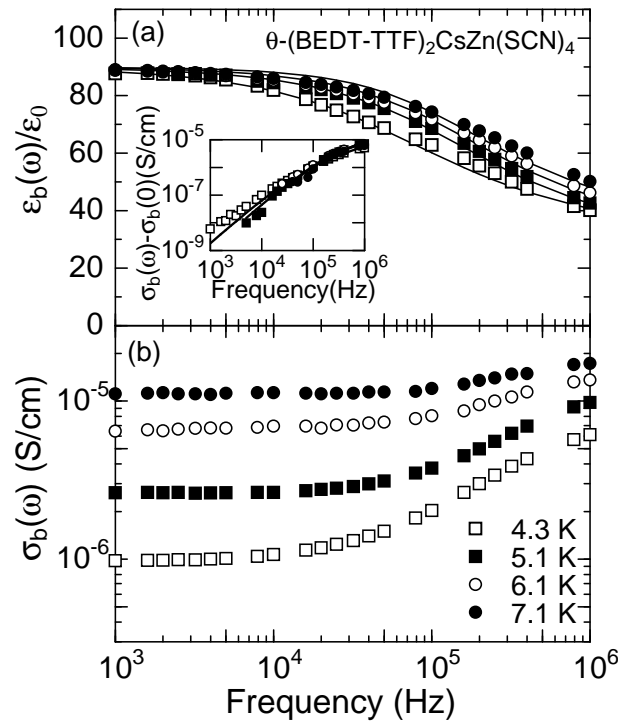}
 \end{center} 
 \caption{
 (a) Dielectric constant and (b) ac conductivity of
 $\theta-$(BEDT-TTF)$_2$CsZn(SCN)$_4$ along the 
 out-of-plane (the $b$ axis) direction
 }
\end{figure}

Figure 1(a) shows $\varepsilon_b(\omega)$ of the $M=$Cs salt.
Reflecting the insulating behavior,
the dielectric response is far from Drude-like:
$\varepsilon_b$ shows a positive sign, and decreases with increasing frequency.
The magnitude is as large as 90 at low frequencies, and such a large 
$\varepsilon$ is rarely observed in conventional insulators.
The large value of 90 is gradually relaxed 
to a small value of 40 at high frequencies,
which can be explained in terms of a generalization of Debye's 
dielectric relaxation (the Havrilliak-Negami formula \cite{HN}) 
given by
\begin{equation}
 \varepsilon (\omega ) = \varepsilon _{\rm HF}+ 
  \frac{\varepsilon_{\rm LF} - \varepsilon_{\rm HF}}
  {[1 +(i \omega \tau )^{1-\alpha }]^{1-\beta }}  
\end{equation}
where $\alpha$ and $\beta$ are dimensionless parameters, and satisfy
$0<\alpha, \beta <1 $.
$\varepsilon_{\rm LF}$ and $\varepsilon_{\rm HF}$ are the 
low- and high-frequency dielectric constants,
and $\tau$ is the relaxation time.
The fitting results are shown by the solid curves in Fig. 1(a),
which satisfactorily fit the measured data.
All the parameters except for $\tau$ are
independent of temperature, which will be discussed 
in the next section.

Figure 1(b) shows $\sigma_b(\omega)$ of the $M=$Cs salt.
As is similar to $\varepsilon_b$,
the frequency dependence suggests relaxation behavior:
A small value of $\sigma$ at low frequencies gradually
increases to a higher value at high frequencies.
Note that $\sigma_b$ has a finite value for $\omega\to0$,
which cannot be ascribed to the dielectric relaxation.
In general, $\sigma$ is expressed as the sum 
of the dc and ac parts given by
\begin{equation}
 \sigma(\omega)=\sigma_{dc}(T)+\sigma_{ac}(\omega,T).
\end{equation}
$\sigma_{dc}$ is derived from the overdamped Drude contribution.
As is well known, the Drude conductivity $ \tilde\sigma_D$ is written by
\begin{equation}
 \tilde\sigma_D = \frac{\varepsilon_0\omega_{pD}^2(\gamma_D+i\omega)}
  {\omega^2+\gamma^2_D},
\end{equation}
where $\omega_{pD}$ and $\gamma_D$ is the plasma frequency and the 
damping rate, respectively.
In the low-frequency limit $\omega\ll \gamma_D$, the above equation 
reduces to 
\begin{eqnarray}
 \sigma_D &=& \varepsilon_0\omega_{pD}^2/\gamma_D =\sigma_{dc}\\
 \varepsilon_D &=& \varepsilon_0(1-\omega_{pD}^2/\gamma^2_D).
\end{eqnarray}
Since $\omega_{pD}^2/\gamma^2_D$ is independent of frequency and small
for the overdamped condition $\gamma_D\gg \omega_{pD}$,
contribution of $\tilde\sigma_D$ to $\varepsilon$ is 
just a small constant shift of the order of unity
to $\varepsilon_{\rm LF}$ and $\varepsilon_{\rm HF}$.
Thus we can safely assume that the contribution of the Drude conductivity 
appears only in the dc limit of the conductivity.
In the inset of Fig. 1(a), 
$\sigma_b(\omega)-\sigma_b(0)$ is fitted with the Havrilliak-Negami formula
with the same parameters as used in $\varepsilon_b(\omega)$ 
through the relation of 
$\sigma_b(\omega)-\sigma_b(0)=-\omega{\rm Im}\varepsilon_b$.
The fitting curves are in excellent agreement with the measured $\sigma_b(\omega)-\sigma_b(0)$,
which consolidates the validity of the Havrilliak-Negami fitting.

\begin{figure}[t]
 \begin{center}
  \includegraphics[width=8cm,clip]{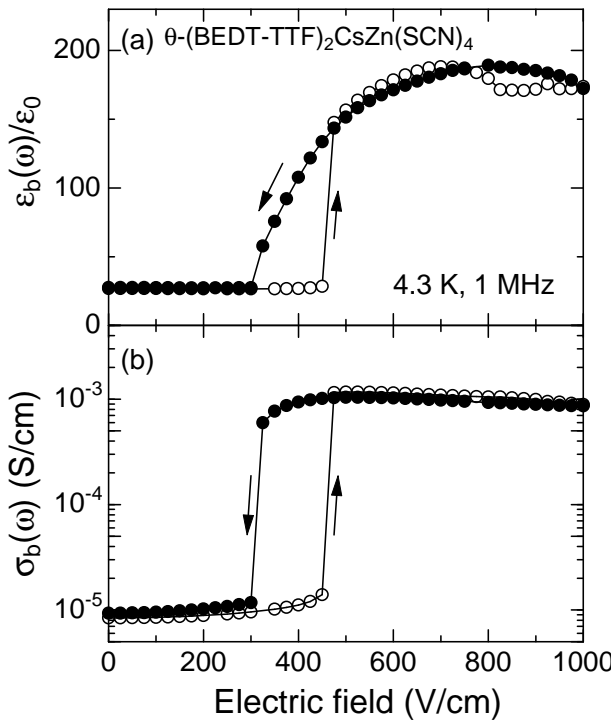}
 \end{center} 
 \caption{
 (a) Dielectric constant and (b) ac conductivity of
 $\theta-$(BEDT-TTF)$_2$CsZn(SCN)$_4$ along the 
 out-of-plane direction as a function of dc bias.
 The measured frequency and temperature are
 1 MHz and 4.2 K, respectively.
 }
\end{figure}

Most unexpectedly, we have found anomalously large bias dependence
of $\varepsilon_b$ and $\sigma_b$, as shown in Fig. 2.
They exhibit a jump to higher values 
at a critical threshold $E_T$ of 450 V/cm with increasing electric field.
The jump is so remarkable that $\sigma_b$ increases by the two orders of magnitude,
and $\varepsilon_b$ also increases up to a large value of 200 simultaneously.
Furthermore, there appears a clear hysteresis with decreasing electric field.

We should emphasize that an artifact such as heating
cannot cause the abrupt jump with the significant hysteresis.
We took the data with the sample immersed in liquid $^4$He,
and carefully evaluated the Joule heating to be a few $\mu$W
at the threshold, which is much smaller than a cooling power 
of liquid $^4$He (typically more than 1W).
We should further note that a preliminary measurement using 
a pulse voltage technique gives identical I-V curves 
to those taken in static dc bias.
The pulse width was changed from 2 to 50 ms with a duration time of 500 ms,
and the voltage was swept from 0 to 10 V.
The jump in the dc conductivity was clearly observed above 6-7 V,
and the conductivity above $E_T$ was independent of the pulse width \cite{sawano}.
This further supports that the heating effect is negligible above $E_T$.

\begin{figure}
 \begin{center}
  \includegraphics[width=8cm,clip]{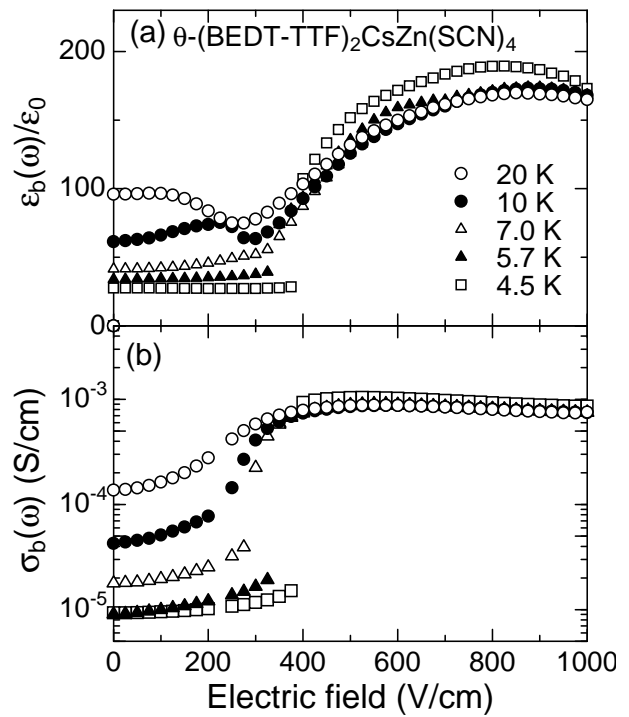}
 \end{center} 
 \caption{
  (a) Dielectric constant and (b) ac conductivity 
 at 1 MHz of
 $\theta-$(BEDT-TTF)$_2$CsZn(SCN)$_4$ along the 
 out-of-plane direction as a function of dc bias.
 The data were measured with increasing electric field.
 }

\end{figure}

The temperature dependence of nonlinear $\varepsilon_b$ 
and $\sigma_b$ is shown in Fig. 3.
The data for increasing electric field are shown,
(the hysteresis in Fig. 2 rapidly disappeared above 4.3 K).
The large bias dependence is clearly observed up to 20 K,
which suggests that the collective excitation survives 
at least up to 20 K.
The threshold field gradually decreases with increasing temperature,
but still remains  finite at 20 K.
Although  $E_T$ is not the phase boundary, it can be a measure for
a boundary between the metallic and insulating states, 
which extends to much higher temperatures.
Another notable feature is that the metallic state (high-bias state)
has essentially temperature-independent $\varepsilon_b$ and $\sigma_b$.
In particular, $\sigma_b$ above $E_T$ is as high as $\sigma_b(0)$ at 50 K.

\begin{figure}[t]
  \begin{center}
  \includegraphics[width=8cm,clip]{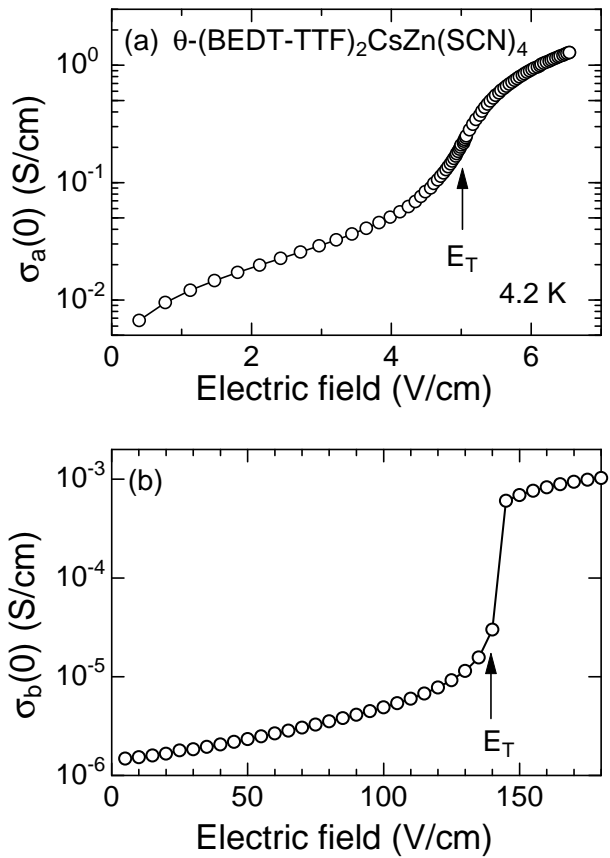}
 \end{center} 
 \caption{
 Nonlinear dc conductivity of $\theta-$(BEDT-TTF)$_2$CsZn(SCN)$_4$
 for (a) the in-plane direction
 and (b) the out-of-plane direction at 4.2 K
 plotted as a function of electric field. The data are measured
 with increasing electric field. }
\end{figure}

In order to examine whether the large bias dependence
exists only along the $b$ direction or not, 
we measured  the non-linear dc conductivity $\sigma(0)$
for the in-plane ($a$ axis) direction as well as along the $b$ direction.
As clearly shown in Fig. 4, $\sigma_a(0)$ and $\sigma_b(0)$
exhibit a large nonlinearity,
although $E_T$ is quite different between the two.
Note that $E_T$ along the $b$ direction
is lower than that in Fig. 2, 
because the sample in Fig. 4 is different from the sample in Figs. 2 and 3.
$E_T$ varies from sample to sample, but
nevertheless the order of the threshold is independent of samples,
which is in the range of 1-10 and 100-1000 V/cm
for the $a$ and $b$ directions, respectively.
Another notable difference is that $\sigma_a(0)$
does not show an abrupt jump, but a steep increase near $E_T$.
It does not show a hysteresis either.

It should be emphasized that the observed giant nonlinear conduction 
cannot be assigned to a single particle excitation.
In general, nonlinear response is observed when the electric energy
gained by the applied electric field $E_{\rm ext}$ 
exceeds thermal energy $k_{\rm B}T$.
Then a characteristic length scale $L$ can be estimated 
from the inquality $eE_{\rm ext}L>k_{\rm B}T$.
In the in-plane conduction, for example, $L$ is estimated to be 
$k_{\rm B}T/eE_T\sim$1 $\mu$m ($T$=4 K and $E_T$=5 V/cm), 
which is much longer than the mean free path of the conduction electron.
It is thus reasonable to assign the giant nonlinear 
conduction to a \textit{collective} excitaion with a 
coherent length scale of 1 $\mu$m.

\section{Discussion}
\subsection{Dielectric response}
As mentioned in the previous section, the large $\varepsilon_b$
of the $M=$Cs salt is expressed by the dielectric relaxation.
A similar $\varepsilon$ is seen in CDW conductors such as 
K$_{0.3}$MoO$_3$\cite{cava},
in which CDW oscillates around pinning centers by
an ac electric field to induce a large dielectric response.
The dielectric relaxation of CDW is phenomenologically derived from 
the overdamped Lorentz oscillator given by
\begin{equation}
 \varepsilon(\omega) = \varepsilon_{\infty} +
  \frac{f}{\omega_0^2-\omega^2 +i\gamma \omega} \qquad (\gamma \gg \omega_0),
\label{lorentz}
\end{equation}
where $f$, $\omega_0$, $\gamma$ are the oscillator strength,
the resonance frequency, and the damping rate, respectively.
Equation (\ref{lorentz}) reduces to the Debye model
of dielectric relaxation in the low frequency limit 
$\omega_0 \gg \omega$ as 
\begin{equation}
 \varepsilon(\omega) = \varepsilon_{\infty} +
  \frac{f}{\omega_0^2} \ \frac{1}{1+i\omega\tau}
  \label{debye}
\end{equation}
where $\tau = \gamma/\omega_0^2$.
In real materials, $\tau$ distributes significantly
owing to many relaxation channels, and $\varepsilon$ obeys 
the Havrilliak-Negami formula given by Eq. (1).

Accordingly $(\varepsilon_{\rm LF}-\varepsilon_{\rm HF})/\varepsilon_0$ is equal to
$f/\omega_0^2$, and thus the large 
$\varepsilon_{\rm LF}-\varepsilon_{\rm HF}$ implies a large $f$ and/or a small $\omega_0$.
In the case of the pinned CDW, $f$ and $\omega_0$ correspond
to the Drude weight ($\varepsilon_0\omega_p^2$) of the order of 1 eV
and the pinning potential of the order of 1 meV,
respectively, which makes $(\omega_p/\omega_0)^2$ as huge as 10$^6$.
A more precise treatment is to solve the equation of motion
for the phase of the CDW order parameter \cite{FLR},
which also gives the overdamped Lorentz oscillator \cite{FLR_e}.

From the viewpoint of charge density modulation,
difference between CDW and CO is very subtle:
the former shows a sinusoidal-wave modulation, 
and the latter shows a square-wave modulation.
In either case, the dielectric response can be described 
in terms of the overdamped Lorentz oscillator.
In fact, charge-ordered materials such as
LuFe$_2$O$_4$ \cite{ikeda,yamada} and 
Pr$_{1-x}$Ca$_x$MnO$_3$ \cite{arima}
show a large $\varepsilon$ described by Eq. (1).
Previously we attributed the large $\varepsilon_b$ of the $M=$Rb salt
to the collective motion of CO \cite{inagaki}.
An important difference from CDW conductors 
is that the dielectric relaxation 
survives in the ``normal'' state far above $T_{\rm MI}$.
This  means that the fluctuation of the pre-formed CO
coexists with the unbound carriers that do not participate in CO.
In this sense, we can say that the electrons in the $M=$Rb salt
are self-organized above $T_{\rm MI}$  to make its density inhomogeneous.

\begin{table}[t]
\caption{\label{tab:table1}
Parameters obtained through fitting to the 
complex dielectric constant by 
a generalization of Debye's 
dielectric relaxation (the Havrilliak-Negami formula)}
\begin{ruledtabular}
\begin{tabular}{lccc}
&Rb ($T>T_{\rm MI}$)&Rb ($T<T_{\rm MI}$)   &Cs\\
\hline
$T$ (K)                            &  200   &  140   & 4.3\\
$\varepsilon_{\rm LF}/\varepsilon_0$   &  38    &  35   & 90\\
$\varepsilon_{\rm HF}/\varepsilon_0$   &  12    &  12   & 29\\
$\alpha$                           &  0.23  & 0.23  & 0.35\\
$\beta$                            &  0     & 0     & 0.15\\
$\tau$  ($\mu$s)                   &  0.17  &  500  & 7\\
$\rho$   (10$^6\Omega$cm)          &  0.016 &  30   & 1\\
\end{tabular}
\end{ruledtabular}
\end{table}

Since the $M=$Cs salt is regarded as a system for $T_{\rm MI}\to0$,
$\varepsilon_b$ in Fig. 1 should be compared with $\varepsilon_b$ of the $M=$Rb salt
above $T_{\rm MI}$.
This picture is consistent with the x-ray diffraction study by Nogami et al. \cite{nogami},
where they succeeded in observing that
a diffuse spot near (0, k, 1/2) grows below 50 K.
This indicates that the short range order of CO 
along the $c^*$ direction grows within the BEDT-TTF layer,
but no three dimensional order is realized down to 4 K.
We should also note that the specific heat \cite{nishio}
and the susceptibility measurements \cite{mori}
did not detect any jump or kink down to 4 K,
implying {\it no} second-order phase  (CDW/SDW/CO) transitions.

The fitting parameters of Eq. (1) are listed in Table 1.
Except for $\tau$ and $\rho$, the obtained parameters are of the same order,
meaning that the dynamics of CO is roughly identical.
In the overdamped Lorentz model, $\tau$ is associated with the damping rate
as $\tau=\gamma/\omega_0^2$.
If the damping rate is determined by the scattering between CO
and the unbound conducting electron, we can expect $\gamma\propto\rho$,
which roughly explains the different magnitudes of $\tau$ and $\rho$
between $M=$Cs and Rb.
We further note that $\varepsilon_{\rm LF}/\varepsilon_0$ 
for $M=$Cs is larger than that for $M=$Rb, 
implying the smaller pinning frequency $\omega_0$ 
as was reported in the impurity-doped CDW material \cite{cava2}.
This is consistent with the large bias dependence seen in Figs. 2-4.
We did not see any significant nonlinear conductions for the $M=$Rb salt.

\begin{figure}[t]
  \begin{center}
  \includegraphics[width=8cm,clip]{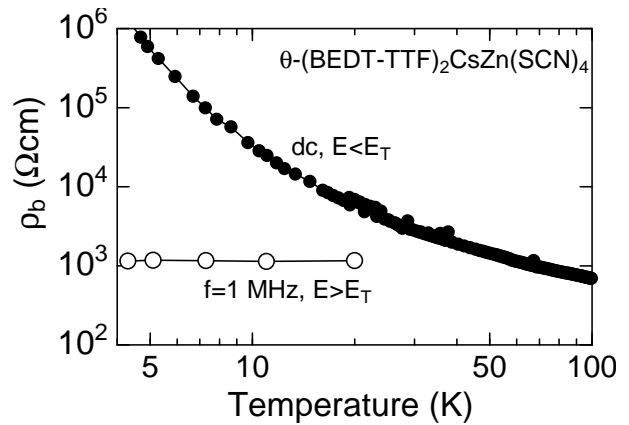}
 \end{center} 
 \caption{
 Out-of-plane resistivities of $\theta-$(BEDT-TTF)$_2$CsZn(SCN)$_4$
 for $E<E_T$ and $E>E_T$
 plotted as a function of temperature.}
\end{figure}

\subsection{Nonlinear conduction}
Although nonlinear conduction is observed in various electron systems,
it is rarely observed in homogeneous bulk materials,
except for the flux flow resistivity of superconductors,
the sliding of CDW/SDW, and the negative resistance in high-mobility semiconductors.
Since the nonlinear conduction in Figs. 2-4 occurs in a low 
electric field of 1-100 V/cm, it should not come from a single
particle excitation, but from a collective excitation.
Actually the data in Fig. 4 is similar 
to the nonlinear conductivity of CDW.

Let us list significant differences from the nonlinear 
conduction of CDW.
(i) The conductivity in Figs. 4(a) and 4(b) is highly 
nonlinear below $E_T$, while
most of the CDW conductors show an ohmic conductivity below $E_T$.
This suggests that the domain sizes and/or the pinning forces
of the pre-formed CO are varied widely in the $M=$Cs salt.
(ii) The CDW conductors show negligibly small nonlinear conduction above $T_c$.
Thus it is quite surprising that the $M=$Cs salt exhibits the huge nonlinearity
of 100-1000 times above $T_{\rm MI}$ ($< 4$ K).
(iii) The observed nonlinearity is much larger than that in 
other two-dimensional conductors \cite{koyano, basletic}.
Large nonlinear conduction has been observed mostly
in one-dimensional CDW conductors such as K$_{0.3}$MoO$_3$,
NbSe$_3$, TaS$_3$\cite{maeda,sneddon}
(iv) The nonlinear conductivity perpendicular to
the charge-density modulation is observed for the first time.

Figure 5 shows the dc resistivity along the $b$ axis
($\rho_b$) below $E_T$ plotted as a function of temperature.
$\rho_b$ increases with decreasing temperature, the
temperature dependence of which is nearly the same 
as that of the in-plane (the $a$ axis) resistivity ($\rho_a$) \cite{mori}.
This indicates that the insulating state is 
more or less three-dimensional.
The low-temperature upturn in $\rho_a$ has been 
understood in terms of variable range hopping (VRH)
---strongly localized state, but
it should be emphasized here that the 
frequency and temperature dependences of $\varepsilon(\omega)$
and $\sigma(\omega)$ seriously contradict the VRH picture.
In Fig. 5, $1/\sigma_b$ (1 MHz) for $E>E_T$ is also plotted as 
a function of temperature, which is nearly the same value of 
$\rho_b$ at 60 K.
This means that the $M=$Cs salt at low temperature
is a ``fake'' insulator that is quite unstable against 
the strong electric field.
Wang et al. observed an anomalous increase in $\sigma_a(\omega)$
of the $M=$Cs salt in the infrared region at low temperatures,
and speculated that the $M=$Cs salt is in the verge of metal-insulator 
transition \cite{wang2}, which might be associated 
with the fake insulating state.

The nonlinear conduction below $E_T$ is usually
weak in the CDW materials. 
A homogeneous CDW order has a single coherent length,
and the corresponding value of $E_T$ is uniquely determined. 
On the contrary, the $M=$Cs salt exhibits 
the nonlinear conduction for $E\ll E_T$ as shown in Fig. 4,
which naturally implies 
a larger scale of the CO domain.
Imagine that the CO domains form a percolative network in a fractal dimension.
Then a larger cluster can move for a smaller electric field,
and the number of the mobile clusters gradually 
increase with increasing field.
This picture is consistent with the charge inhomogeneity in the 
high-temperature superconductor \cite{davis}.

The most puzzling finding is the large hysteresis of
$\varepsilon_b$ and $\sigma_b$ shown in Fig. 2.
At first glace, this resembles the switching behavior 
of K$_{0.3}$MoO$_3$ \cite{maeda}, which was phenomenologically
understood in terms of the damping rate depending 
on the CDW sliding velocity \cite{littlewood}.
However, the electric field was \textit{perpendicular} to the charge modulation 
in the present case, and we cannot expect the sliding of CO across the layers.
We cannot expect either that a CO domain 
hops from one layer to other, because it requires enormous numbers
of electrons to tunnel simultaneously.
We should further note that $\varepsilon$ above $E_T$ decreases 
with increasing bias voltage for conventional CDW conductors \cite{sneddon},
which is seriously incompatible with the data in Fig. 2(a).

Finally we will append a note that
we have found a similar bias dependence in the dc conductivity
of $\theta-$(BEDT-TTF)$_2$CsCo(SCN)$_4$,
which has nearly the same structure as $\theta-$(BEDT-TTF)$_2$CsZn(SCN)$_4$.
This strongly suggests that 
the giant nonlinear conduction is 
a generic feature for some $\theta$-type salts \cite{sawano}.

\begin{figure}[t]
 \begin{center}
  \includegraphics[width=8cm,clip]{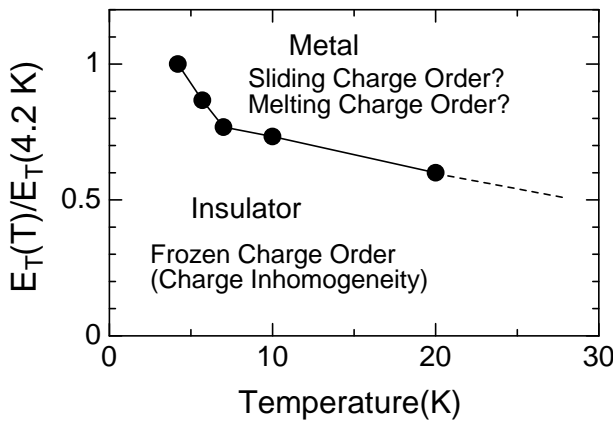}
 \end{center} 
 \caption{
 Electronic pseudo-phase diagram for 
 $\theta-$(BEDT-TTF)$_2$CsZn(SCN)$_4$
 }
\end{figure}

\subsection{Inhomogeneity in a homogeneous system}
As mentioned in the Introduction, the organic salts are
one of the cleanest solids currently available, in which 
point defects, dislocations and solid solutions are expected 
to be negligibly small.
Nevertheless, $\rho_a$ for $M=$Cs is 30 m$\Omega$cm at 300 K \cite{mori},
which is indeed anomalously high.
In a two-dimensional conductor,
the mean free path $\ell$ is directly evaluated 
from the in-plane resistivity through the relation 
\begin{equation}
 k_F\ell = {hc_0}/{e^2\rho_a},
\end{equation}
where $c_0$ is the interlayer spacing \cite{ando}.
Then 30 m$\Omega$cm gives $k_F\ell=$0.1,
and such a short $\ell$ is difficult to understand in the clean system.
The metallic behavior ($d\rho_a/dT>0$) at 300 K
is also difficult to understand, because de Broglie wavelength
of the conducting electron is much shorter than $\ell$ for
$k_F\ell\ll1$.

We can clear up these difficulties, if the 
electronic states of the $M=$Cs salt are spacially inhomogeneous,
because Eq. (8) is derived for a homogenous system.
Suppose most of the conduction electrons are nearly frozen to
form CO domains, and the unbound electrons
flow in a percolation network consisting of the domain boundaries.
Then the effective cross section for the current path is
much smaller than the bulk cross section of the sample, 
and the ``apparent'' resistivity would be much higher.
If $k_F\ell$ is still larger than unity in the current path,
the conduction can be metallic to cause $d\rho/dT>0$.
Judging from the anomalously high in-plane resistivity, 
we think that the self-organization  of the electrons 
in $M=$Cs readily starts even at room temperature.
Then the self-organized CO is gradually 
frozen with decreasing temperature,
and eventually shows the CDW-like collective excitation.

Let us discuss the reason why the self-organized inhomogeneity 
is realized in the $\theta-$type salts.
Mori \cite{takehiko} calculated the ground-state energy for the $\theta-$type salts
using the extended Hubbard model, and discussed 
possible CO patterns at finite temperatures.
He found that the on- and off-site Coulomb repulsions
are of the same order, and their delicate balance causes various CO patterns.
In particular, ``horizontal'', ``diagonal'' and ``3-fold'' patterns
are nearly degenerate in the $M=$Cs salt, and
thus it is highly probable that different charge orders
coexist.
Furthermore, these charge orders would be strongly coupled 
with phonons, as was evidenced by the strong Fano effects 
in the infrared spectroscopy \cite{wang,wang2}.
Thus CO domains of different patterns store 
finite elastic energies, which prevents the domains  to grow in size \cite{relaxor1}.
As a result, homogeneous phase transition is seriously suppressed, 
and the domains are frozen at low temperatures
like polar nano-regions in relaxor ferroelectrics \cite{relaxor2}.
We therefore expect that the $M=$Cs salt 
can be a canonical material showing the intrinsic inhomogeneity
in the electron density.

A remaining issue is up to what temperature
the collective excitation can be observed.
Figure 6 shows $E_T$ in Fig. 3 (b) as a function of temperature.
Clearly, the phase boundary would exist above 20 K.
If we extrapolate the data to higher temperatures 
as indicated by the dotted line, 
we can estimate the temperature for $E_T=0$ to be 65 K,
which is roughly the same temperature at which
$1/\sigma_b({\rm 1 MHz})$ in the high-bias state is equal to 
$\rho_b$ in Fig. 5.
To measure $E_T$ at higher temperatures, we should employ 
a pulse generator to avoid the Joule heating,
which will be tried in future work.

The metallic state above $E>E_T$ is not yet understood.
In CDW conductors, the high-biased state is the state
that CDW is sliding with a finite damping from
the normal electrons, which has been established 
from the diffraction experiment in high electric fields.
In the present case, the $M=$Cs salt only has the short range order
of CO, where the phase of CO acquires no rigidity.
Then the sliding of CO is difficult to expect.
Another choice is the melting of CO by electric field,
like the vortex lattice melting by magnetic field 
in a type-II superconductor.
To clarify this problem, diffraction measurements
in electric fields are indispensable,
and microscopic theory to explain our findings
is also necessary.

\section{Summary}
We have observed the large dielectric response and giant nonlinear 
conduction in $\theta-$(BEDT-TTF)$_2$CsZn(SCN)$_4$.
The data resemble those for charge-density-wave (CDW) conductors 
such as K$_{0.3}$MoO$_3$.
However, $\theta-$(BEDT-TTF)$_2$CsZn(SCN)$_4$ shows no second-order transition
down to 4 K, which implies that it exhibits CDW-like collective excitations
in the ``normal'' state.
Therefore we expect that this material can belong to a new hierarchy
of condensed matter, in which conduction electrons are self-organized
to have collective modes without long range order.

\section*{Acknowledgments}
The authors would like to thank W. Kobayashi, S. Kurihara, A. Maeda, H. Matsukawa, J. Goryo,
H. Yoshimoto, K. Kanoda, M. Ogata and H. Fukuyama 
for fruitful discussion and useful advices.
They also appreciate F. Sawano for collaboration, and Y. Yoshino for technical support.
This work was partially supported by MEXT, the Grant-in-Aid for 
Scientific Research (C), 2001, no.13640374,
and by the Grant-in-Aid for The 21st Century COE Program (Holistic
Research and Education Center for Physics Self-organization Systems) at
Waseda University.

\end{document}